\documentclass[letterpaper,12pt,copyedit]{article}   
\usepackage{osajnl2} 
\usepackage[draft]{hyperref} 
\usepackage{endfloat}
\usepackage{xspace}
\usepackage{graphicx}
\usepackage{tabularx}
\newcommand{\ignore}[1]{}
\providecommand{\ao}{}
\renewcommand{\ao}{adaptive optics (AO)\renewcommand{\ao}{AO\xspace}\xspace}
\newcommand{\wfs}{wavefront sensor (WFS)\renewcommand{\wfs}{WFS\xspace}\renewcommand{\wfss}{WFSs\xspace}\xspace}
\newcommand{\wfss}{wavefront sensors (WFSs)\renewcommand{\wfs}{WFS\xspace}\renewcommand{\wfss}{WFSs\xspace}\xspace}
\newcommand{\shwfs}{Shack-Hartmann \wfs (SHWFS)\renewcommand{\shwfs}{SHWFS\xspace}\xspace}
\newcommand{\dm}{deformable mirror (DM)\renewcommand{\dm}{DM\xspace}\renewcommand{\dms}{DMs\xspace}\renewcommand{\Dms}{DMs\xspace}\renewcommand{\Dm}{DM\xspace}\xspace}
\newcommand{\dms}{deformable mirrors (DMs)\renewcommand{\dm}{DM\xspace}\renewcommand{\dms}{DMs\xspace}\renewcommand{\Dms}{DMs\xspace}\renewcommand{\Dm}{DM\xspace}\xspace}
\newcommand{\Dms}{Deformable mirrors (DMs)\renewcommand{\dm}{DM\xspace}\renewcommand{\dms}{DMs\xspace}\renewcommand{\Dms}{DMs\xspace}\renewcommand{\Dm}{DM\xspace}\xspace}
\newcommand{\Dm}{Deformable mirror (DM)\renewcommand{\dm}{DM\xspace}\renewcommand{\dms}{DMs\xspace}\renewcommand{\Dms}{DMs\xspace}\renewcommand{\Dm}{DM\xspace}\xspace}

\newcommand{\shs}{Shack-Hartmann sensor (SHS)\renewcommand{\shs}{SHS\xspace}\renewcommand{\shss}{SHSs\xspace}\xspace}
\newcommand{\shss}{Shack-Hartmann sensors (SHSs)\renewcommand{\shs}{SHS\xspace}\renewcommand{\shss}{SHSs\xspace}\xspace}
\newcommand{\lgs}{laser guide star (LGS)\renewcommand{\lgs}{LGS\xspace}\renewcommand{\lgss}{LGSs\xspace}\xspace}
\newcommand{\lgss}{laser guide stars (LGSs)\renewcommand{\lgs}{LGS\xspace}\renewcommand{\lgss}{LGSs\xspace}\xspace}
\newcommand{\ngs}{natural guide star (NGS)\renewcommand{\ngs}{NGS\xspace}\renewcommand{\ngss}{NGSs\xspace}\xspace}
\newcommand{\ngss}{natural guide stars (NGSs)\renewcommand{\ngs}{NGS\xspace}\renewcommand{\ngss}{NGSs\xspace}\xspace}
\newcommand{\mems}{Micro-Electro-Mechanical Systems (MEMS)\renewcommand{\mems}{MEMS\xspace}\xspace}
\newcommand{\snr}{signal to noise ratio (SNR)\renewcommand{\snr}{SNR\xspace}\xspace}
\newcommand{\moao}{multi-object \ao (MOAO)\renewcommand{\moao}{MOAO\xspace}\xspace}
\newcommand{\mcao}{multi-conjugate adaptive optics (MCAO)\renewcommand{\mcao}{MCAO\xspace}\xspace}
\newcommand{\ltao}{laser tomographic adaptive optics (LTAO)\renewcommand{\ltao}{LTAO\xspace}\xspace}
\newcommand{\cpu}{central processing unit (CPU)\renewcommand{\cpu}{CPU\xspace}\renewcommand{\cpus}{CPUs\xspace}\xspace}
\newcommand{\cpus}{central processing units (CPUs)\renewcommand{\cpu}{CPU\xspace}\renewcommand{\cpus}{CPUs\xspace}\xspace}
\newcommand{\psf}{point spread function (PSF)\renewcommand{\psf}{PSF\xspace}\renewcommand{\psfs}{PSFs\xspace}\xspace}
\newcommand{\psfs}{point spread functions (PSFs)\renewcommand{\psf}{PSF\xspace}\renewcommand{\psfs}{PSFs\xspace}\xspace}
\newcommand{\fpga}{field programmable gate array (FPGA)\renewcommand{\fpga}{FPGA\xspace}\renewcommand{\fpgas}{FPGAs\xspace}\xspace}
\newcommand{\fpgas}{field programmable gate arrays (FPGAs)\renewcommand{\fpga}{FPGA\xspace}\renewcommand{\fpgas}{FPGAs\xspace}\xspace}
\newcommand{\sor}{successive over-relaxation (SOR)\renewcommand{\sor}{SOR\xspace}\xspace}
\newcommand{\fdpcg}{Fourier domain pre-conditioned gradient (FDPCG)\renewcommand{\fdpcg}{FDPCG\xspace}\xspace}
\newcommand{\map}{maximum a-posteriori (MAP)\renewcommand{\map}{MAP\xspace}\xspace}
\newcommand{\elt}{Extremely Large Telescope (ELT)\renewcommand{\elt}{ELT\xspace}\renewcommand{\elts}{ELTs\xspace}\xspace}
\newcommand{\elts}{Extremely Large Telescopes (ELTs)\renewcommand{\elt}{ELT\xspace}\renewcommand{\elts}{ELTs\xspace}\xspace}
\newcommand{\dugall}{Durham University generalised adaptive optics laser laboratory (DUGALL)\renewcommand{\dugall}{DUGALL\xspace}\xspace}
\newcommand{\fwhm}{full-width at half-maximum (FWHM)\renewcommand{\fwhm}{FWHM\xspace}\xspace}
\newcommand{\wht}{William Herschel Telescope (WHT)\renewcommand{\wht}{WHT\xspace}\xspace}
\newcommand{\emccd}{electron multiplying CCD (EMCCD)\renewcommand{\emccd}{EMCCD\xspace}\xspace}
\newcommand{\dasp}{Durham \ao simulation platform (DASP)\renewcommand{\dasp}{DASP\xspace}\xspace}
\newcommand{\eelt}{European \elt (E-ELT)\renewcommand{\eelt}{E-ELT\xspace}\xspace}
\newcommand{\mpi}{Message Passing Interface (MPI)\renewcommand{\mpi}{MPI\xspace}\xspace}
\newcommand{\smp}{symmetric multi-processing (SMP)\renewcommand{\smp}{SMP\xspace}\xspace}
\newcommand{\svd}{singular value decomposition (SVD)\renewcommand{\svd}{SVD\xspace}\xspace}
\newcommand{\gpu}{graphical processing unit (GPU)\renewcommand{\gpu}{GPU\xspace}\renewcommand{\gpus}{GPUs\xspace}\xspace}
\newcommand{\gpus}{graphical processing units (GPUs)\renewcommand{\gpu}{GPU\xspace}\renewcommand{\gpus}{GPUs\xspace}\xspace}
\newcommand{\fft}{fast Fourier transform (FFT)\renewcommand{\fft}{FFT\xspace}\xspace}
\newcommand{\ifu}{integral field unit (IFU)\renewcommand{\ifu}{IFU\xspace}\xspace}
\newcommand{\darc}{the Durham \ao real-time controller (DARC)\renewcommand{\darc}{DARC\xspace}\renewcommand{\Darc}{DARC\xspace}\xspace}
\newcommand{\Darc}{The Durham \ao real-time controller (DARC)\renewcommand{\darc}{DARC\xspace}\renewcommand{\Darc}{DARC\xspace}\xspace}
\newcommand{\cots}{commercial off-the-shelf (COTS)\renewcommand{\cots}{COTS\xspace}\xspace}
\newcommand{\rtcp}{real-time control pipeline (RTCP)\renewcommand{\rtcp}{RTCP\xspace}\xspace}
\newcommand{\rms}{root-mean-square (RMS)\renewcommand{\rms}{RMS\xspace}\xspace}
\newcommand{\sFPDP}{serial Front Panel Data Port (sFPDP)\renewcommand{\sFPDP}{sFPDP\xspace}\xspace}
\newcommand{\tmt}{Thirty Meter Telescope (TMT)\renewcommand{\tmt}{TMT\xspace}\xspace}

\begin{document}

\title{Considerations for EAGLE from Monte-Carlo adaptive optics simulation}

\author{Alastair Basden,$^{1,*}$ Richard Myers,$^1$ and Timothy Butterley$^{1}$}
\address{$^1$Department of Physics, Durham University, South Road,
  Durham DH1 3LE, UK}
\address{$^*$Corresponding author: a.g.basden@durham.ac.uk}

\begin{abstract}
The EAGLE instrument for the E-ELT is a multi-IFU spectrograph, that
uses a MOAO system for wavefront correction of interesting lines of
sight.  We present a Monte-Carlo AO simulation package that has been
used to model the performace of EAGLE, and provide results, including
comparisons with an analytical code.  These results include an
investigation of the performance of compressed reconstructor
representations that have the potential to significantly reduce the
complexity of a real-time control system when implemented.
\end{abstract}


\maketitle 

\section{Introduction}
The next generation of optical ground-based \elts is currently in
the design phase, with plans for primary mirror diameters of over 30~m
\cite{tmt,eelt}.  Once built, these facilities will allow
astronomers to probe the universe with unprecedented sensitivity and
very high resolution.  A suite of instruments for these telescopes is
planned, allowing many different observation goals to be met.

The EAGLE instrument for the planned 42~m \eelt is currently in the design phase
\cite{eagleScience}.  It is a multi-object integral field unit
spectrograph using \ao with a \moao system to correct
incoming wavefronts in open-loop, using wavefront sensors which do not
sense the corrections made to the science fields.  Baseline designs
for EAGLE include up to 11 wavefront sensors, using laser and natural
guide stars.  It is envisaged that there will be 20 science field
pick-offs, allowing good \ao correction in 20 separate fields, each
1.5~arcseconds diameter, simultaneously across a five arc-minute
field.  The use of a multi-object \ao system allows good atmospheric
correction to be achieved for selected objects across a wide field of
view.  

Part of the design phase for EAGLE includes extensive simulation and
modelling of the \ao performance, since \ao is an essential
part of the instrument design.  The simulation and modelling are
carried out in two phases.  First, an analytical code is used to
obtain an order-of-magnitude performance estimate, covering a large
parameter space relatively quickly.  However, many fine details which
are essential to include for \elt scale designs are not included.  A
Monte-Carlo code is then used to fill in details giving a more
reliable performance estimate, including non-linear effects, and noise
sources.  However Monte-Carlo simulation has far greater computational
requirements, so a reduced parameter space is considered.  

The \dasp is a Monte-Carlo code which can be used for the simulation
of any common form of \ao system (including classical \ao, laser
tomographic \ao, multi-conjugate \ao and \moao
\cite{2007SPIE.6584E...6D}), and has been developed specifically with
\elt simulation in mind \cite{basden5}.  It is an end-to-end
time-domain code and is parallelised, allowing it to be used across a
computing cluster using the \mpi library to reduce computation time.
It includes detailed models of telescope and \ao systems, allowing
high fidelity models to be produced.

The development of a real-time control system for EAGLE is a
challenge.  There are expected to be of order $10^5$ wavefront slope
measurements, and of order $10^4$ \dm actuators to control per science
path (of which there will be about 20).  These wavefront slope
measurements will be used to update the \dm actuators at about 250~Hz.
It is likely that EAGLE will use a conventional matrix-vector based
wavefront reconstruction, though other techniques, such as iterative
algorithms have not been ruled out.  In this paper, we concentrate
only on the matrix-vector based wavefront reconstruction and consider
some details that may make this easier to implement in hardware.  In
order to access all the elements of the control matrix for each new
set of slope measurements, a data rate of order 20~TBs$^{-1}$ is
therefore required (assuming four bytes per matrix element).  This
will require a very advanced control system, and so any achievable
simplifications are desirable.

Here, we present some recent results obtained from the simulation of
EAGLE using the \dasp.  Some of these results are compared with those
from an analytical code, produced independently by another member of
the EAGLE consortium, where appropriate.  However, the analytical code
is unable to include non-linear effects, and so can only be used for
rough performance estimates.  The simulations presented here include
an investigation of compressed wavefront reconstructor algorithms
which could simplify real-time control system design.  The technical
difficulty of the real-time control system design is such that it
should be considered even at the early design phases.  We also discuss
the issue of Shack-Hartmann sensor non-linearities for open-loop
systems.  

In \S2, we describe the simulations that have been carried out, in \S3
we give results, and conclusions are made in \S4.

\section{Simulation description}
There are several possible designs for EAGLE, with different \lgs and
\ngs requirements, based on trade-offs between cost, performance and
sky coverage.  Here, we concentrate on a design with nine \lgss
equally spaced around a ring with a 7.3~arcmin diameter and a single
\ngs with $16\times16$ sub-apertures used for low order mode
correction (tip, tilt, focus and astigmatism) which \lgs sensors are
easily measure usefully.  We assume that the \lgss are centre-launched
and have an elongation of 5~arcsec at the edge of the telescope pupil
(maximum elongation).  No measures to mitigate this elongation are
made and wavefront slope computation uses a centre of gravity
algorithm.  The telescope diameter is assumed to be 42~m, and the
\wfss have $84\times84$ sub-apertures each with $20\times20$ pixels,
unless otherwise stated, requiring a \wfs detector with
$1680\times1680$ pixels.  Each science field pick-off has its own
deformable mirror ($85\times85$ actuators) with wavefront control
optimised along the line-of-sight for this field.  Unless otherwise
stated, the results presented here are for a target at the centre of
the field, i.e.\ in the middle of the \lgs ring.  The science
wavelength is H-band (1.65~$\mu$m).  We use a virtual \dm formulation
for wavefront control, placing virtual \dms conjugate to the height of
strong turbulence.  This allows us to reconstruct the atmospheric
turbulence at the positions of these virtual \dms, and use this
knowledge to determine the shape that should be given to the physical
\dm for this science field.  In the simulations here, the shape given
to each \moao physical \dm is the sum of the virtual \dms projected
along the line-of-sight for this science field.  The virtual wavefront
reconstruction is performed using a standard truncated least-squares
matrix-vector algorithm with the vector containing the latest
wavefront slope measurements, and the matrix being the pseudo-inverse
of the system interaction matrix (the measured \wfs response to
perturbations induced on the virtual \dms).  The simulations presented
here all assume two discrete layers of atmospheric turbulence, and two
virtual \dms unless otherwise stated.  It should be noted that the
estimated performance reported here should be seen as optimistic due
to the simple nature of a two layer profile.  The atmospheric outer
scale is taken as 50~m, and Fried's parameter is 10.6~cm (at 500~nm),
corresponding to a seeing of 0.95~arcsec.  These values are as used in
simulations carried out by Fusco et.\ al.\ for the EAGLE consortium
\cite{fuscoEagle} which our Monte-Carlo simulations are used to
verify.  The update rate of the \ao loop is 250~Hz, and unless
otherwise stated, a delay (latency) of 4~ms between wavefront
detection and correction is simulated.

The deformable mirrors are operated in open-loop, i.e.\ the \wfss do
not sense changes made to the \dms.  

In these simulations, atmospheric phase screens are translated across
the telescope pupil assuming a frozen-flow turbulence model
\cite{tatarski}.  The sections of these screens relevant to a given
line-of-sight at a given time are then selected (with sub-pixel
interpolation) and summed (with interpolation for a source at finite
distance, e.g.\ a \lgs).  These line-of-sight pupil phases are then
used as input to Shack-Hartmann \wfs models, which produce a simulated
noisy Shack-Hartmann image, and to generate science camera images,
before and after correction of the phase using a \dm.  The \dms are
controlled by a wavefront reconstructor, which uses the slope
measurements taken by the \wfss to compute the correction to be
applied.  We use a centre of gravity algorithm for wavefront sensing.
Sodium laser spots (as produced by a \lgs) are assumed to form at
90~km with a depth of about 10~km and a Gaussian distribution.  The
performance of the AO system can be measured as a function of time,
and the average long-exposure performance is also obtained.  These
simulations include many noise sources including detector noise,
photon shot noise, laser guide star elongation, and \wfs
non-linearities.  The simulation code is therefore suited to the high
fidelity modelling of \ao systems.

\subsection{Parameter space}
We have covered a large parameter space during these simulations, and
with our available hardware we are able to cover about 7 parameter
points per day including generation of interaction and control matrices.
Parameters that have been explored include:
\begin{enumerate}
\item Deformable mirror mis-conjugation
\item Control matrix representation (investigating reductions in
  control matrix size to simplify real-time control system development)
\item \wfs linearisation
\item Deformable mirror mis-alignment
\item Wavefront sensor pixel scale
\item Zenith angle
\item \lgs power
\item \wfs read-out noise
\item Secondary mirror support obscuration
\item Woofer-tweeter configuration
\item Sodium layer profile
\end{enumerate}

Here, we consider further the first three parameters.  \Dm
mis-conjugation can occur when knowledge of the atmosphere is not
perfect, and introduces an additional error into the correction
applied to the \dm.  Control matrix representation is an important
consideration for the design and development of real-time control
systems, allowing designs to be simplified and costs reduced when the
memory required to store a control matrix is reduced.  \wfs
linearisation is necessary for open-loop systems because
Shack-Hartmann based \wfss have a slightly non-linear response to
incident wavefront slope, and attempts to calibrate and correct this
non-linearity can improve \ao performance.

\section{Results}

\subsection{Correction across the field of view}
\begin{figure}
\includegraphics[width=7cm]{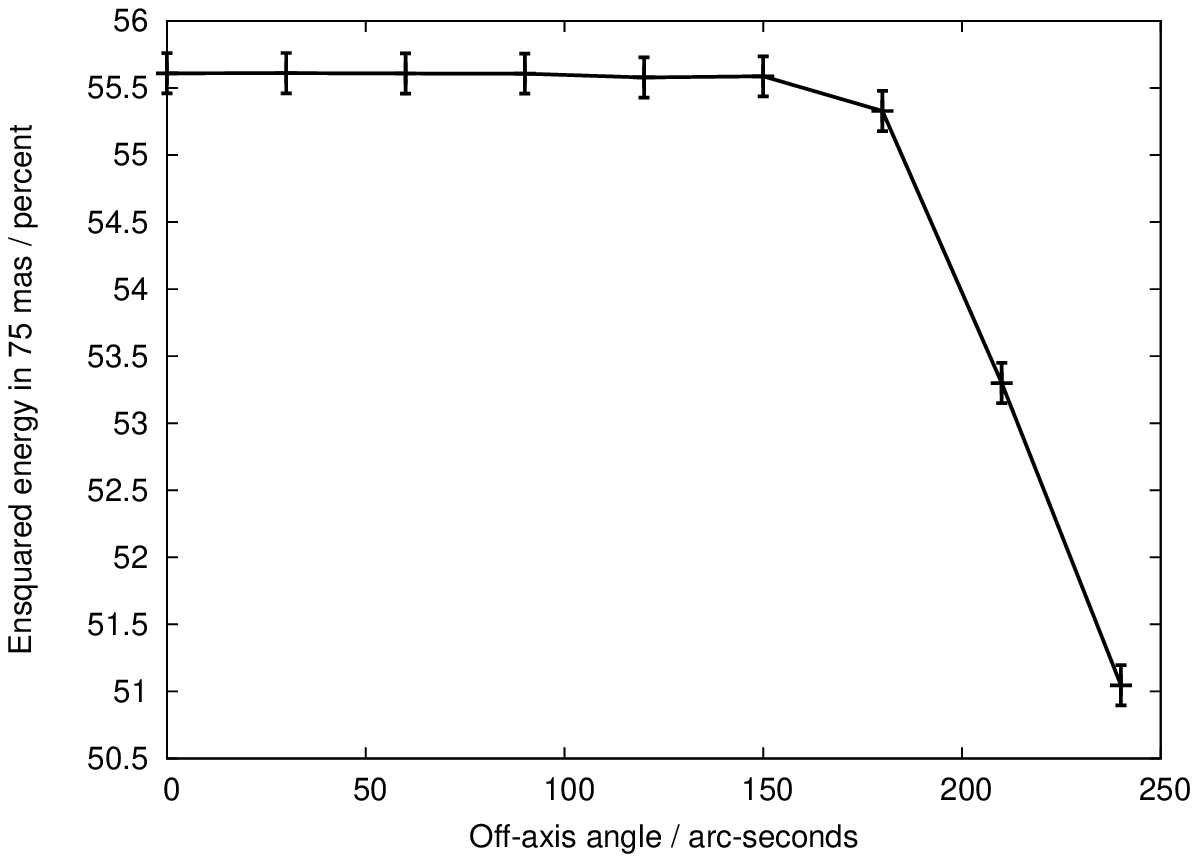}
\caption{A figure showing AO performance for different line-of-sight
  directions across the field of view.  The centre of the field of
  view is at 0~arcsec, and an LGS is at 219~arcsec.}
\label{fig:nsci}
\end{figure}

Since \moao systems operate in open-loop (the wavefront sensors do not
sense the applied wavefront corrections), the corrections applied to
the wavefront can be made along any line-of-sight.  Unless stated
otherwise, results presented here are for a line-of-sight at the
centre of the field of view, i.e.\ the direction corresponding to the
centre of the \lgs ring.  However, it is instructive to compare
expected performance across the field of view, and Fig.~\ref{fig:nsci}
shows performance as a function of position across the field.  In this
figure, the corrected line-of-sight is moved from the on-axis location
in a direction towards and past one of the \lgss (at 219~arcseconds).
It can be seen that correction is uniform for most of the field of
view within the \lgs ring, and performance falls once the
line-of-sight is close to the \lgs ring, due to poor sampling of
turbulence at these locations; turbulence here is only sampled by one
\wfs and so cannot be reconstructed well, while turbulence in the
centre of the field of view is sampled by many \wfss.

\subsection{DM mis-conjugation}
A multi-conjugate \ao system (including a \moao system such as EAGLE,
using virtual \dms) requires information about the strength and
position of turbulent atmospheric layers to operate most effectively.
\Dms are then conjugated at the locations of the most dominant layers.
However, if mis-conjugation occurs, for example because layer
positions are not well known, the \ao system performance will be
degraded.  

\begin{figure}
\includegraphics[width=7cm]{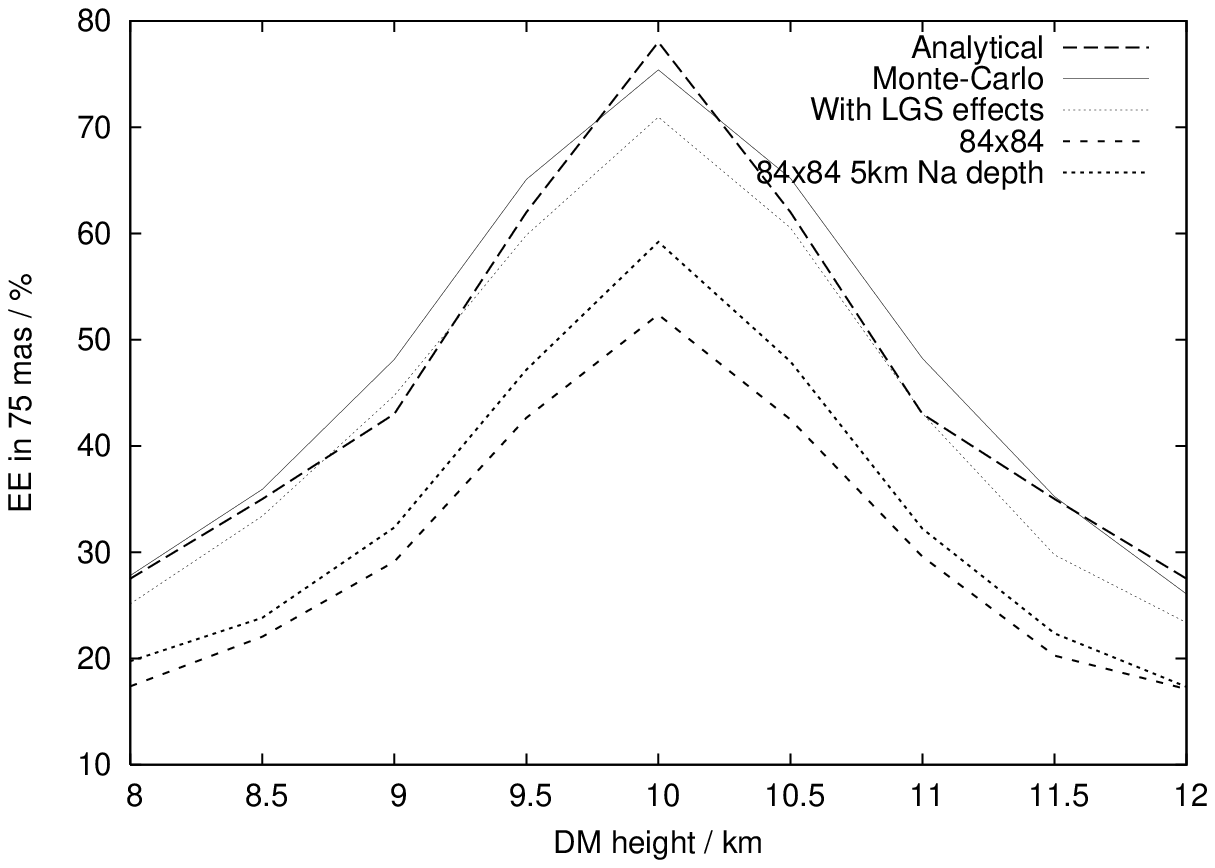}
\caption{A figure showing the effect of DM mis-conjgation on AO system
  performance, which is represented by ensquared energy falling in a
  75~mas diameter box with a science wavelength of 1.65 microns.  The
  long-dashed curve shows analytical results for a $110\times110$
  sub-aperture system, with the solid curve showing the Monte-Carlo
  simulation equivalent.  The dotted curve just below this shows the
  performance reduction when a more realistic simulation including
  cone effect and LGS spot elongation is included.  The two lowest
  curves show the performance reduction when the sub-aperture order is
  reduced to $84\times84$, with two different sodium layer
  profiles. misconjugationF1.eps}
\label{fig:misconjugation}
\end{figure}

Fig.~\ref{fig:misconjugation} shows how the performance of an \ao
system (correcting at H-band) is degraded by mis-conjugation.  Here,
dominant turbulent layers were placed at 0~km and 10~km, and two
virtual DMs placed at 0~km, and at a varying height between 8--12~km.
This figure demonstrates that it is necessary to be able to conjugate
\dms to within a few hundred meters of dominant turbulence.
Analytical results provided by Fusco et al~\cite{fuscoEagle}, which
replace the \lgss with \ngss (the analytical code cannot model cone
effect or spot elongation) for a
system with $110\times110$ sub-apertures per wavefront sensor (long
dashes) are shown to be slightly optimistic for perfect conjugation
when compared with equivalent Monte-Carlo results (solid curve), with
$110\times110$ sub-apertures (for comparison purposes) for each
wavefront sensor.  When a cone effect due to the laser spot being at a
finite distance (meaning only a cone of turbulence is sampled by the
\wfs) and spot elongation caused by the three-dimensional nature of
sodium emission are included in the simulations (dotted curve),
performance is seen to fall by about ten percent.  These effects are
not modelled in the analytical results.  A reduction of \wfs order to
$84\times84$ sub-apertures is shown to further reduce performance (the
lower two curves).  Performance is also shown to be dependent on the
sodium profile.

It should be noted that these simulations are a simplification of the
true situation where there will be many more turbulent layers, each
with finite thickness.  However, for all of these cases, the general
trend with mis-conjugation is clear, implying that a \dm should be
conjugated to dominant turbulence with an accuracy of a few hundred
meters.  This places constraints on the design of turbulence profiling
systems.  

\begin{figure}
\includegraphics[width=7cm]{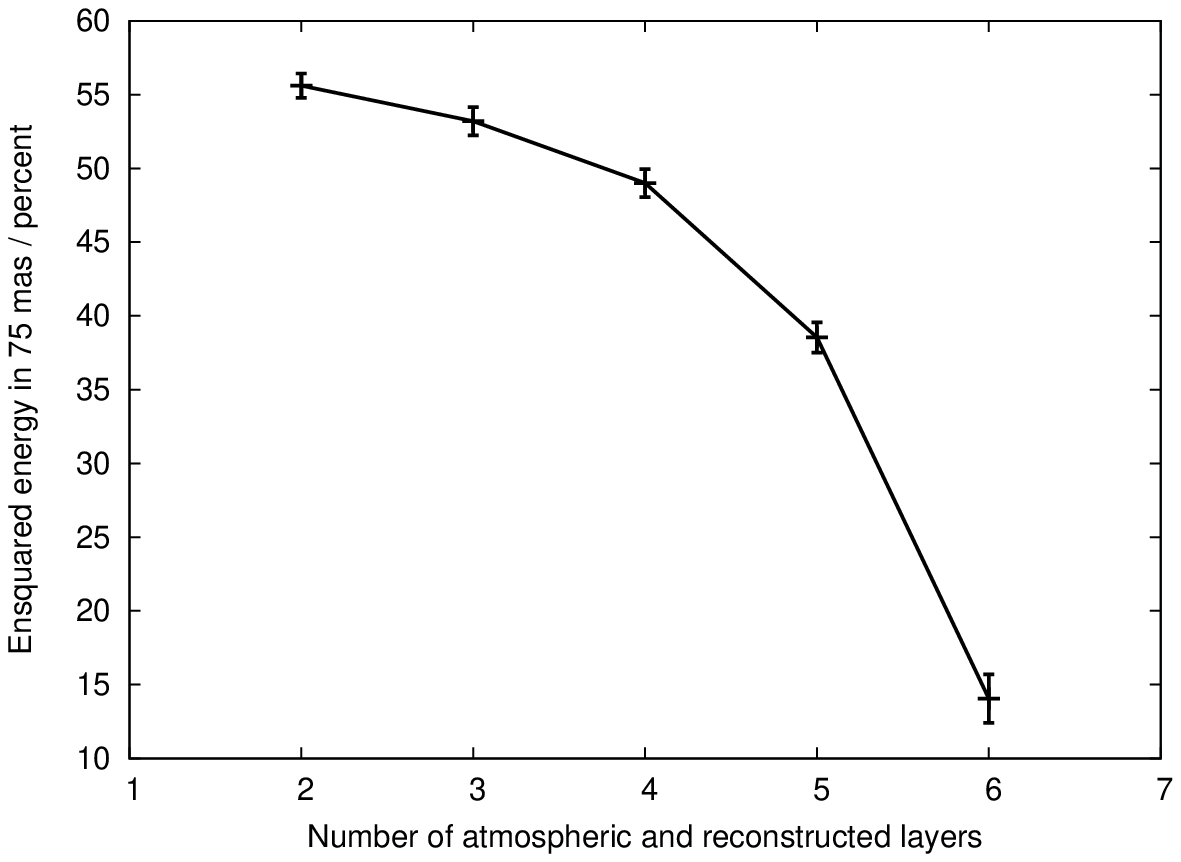}
\caption{A figure showing AO system performance as a function of
  number of atmospheric layers and virtual DMs simulated.}
\label{fig:nlayer}
\end{figure}

Fig.~\ref{fig:nlayer} shows how the simulated \ao system performance
falls as a function of number of atmospheric layers in these
simulations.  Here, we have not sought to optimise the wavefront
reconstruction in any way, using a simple truncated least-squares
wavefront reconstructor.  A virtual \dm has been placed conjugate to
each layer, with an actuator spacing calculated to minimise fitting
error, following \cite{gavelDMFittingError}.  The ideal number of
virtual \dms, their conjugate heights and the actuator spacings to use
to optimise \moao system performance is a subject of on-going
research.  Here, we do not consider the effect of \dm mis-conjugation
when there are more than two atmospheric layers.
Table~\ref{table:atmos} shows the parameters used for these multiple
layer simulations, as provided by Fusco et.\ al.\ \cite{fuscoEagle}.
A global Fried parameter of 10.6~cm and an outer scale of 50~m were
used.  It should be noted that wavefront reconstruction uses a least
squares algorithm.  The use of a minimum variance wavefront
reconstruction may improve performance.  However, this shows that the
performance of EAGLE is likely to fall when the atmospheric turbulence
is heavily layered.

\begin{figure}
\begin{tabularx}{\linewidth}{cXX}
Number of layer & Layer heights & Layer strengths\\ \hline
2 & 0, 12800 & 0.92, 0.08\\
3 & 0, 1800, 12800 & 0.77, 0.17, 0.06\\
4 & 0, 1800, 4500, 12800 & 0.67, 0.15, 0.13, 0.05\\
5 & 0, 300, 1800, 4500, 12800 & 0.53, 0.21, 0.12, 0.10, 0.045\\
6 & 0, 300, 900, 1800, 4500, 12800 & 0.47, 0.18, 0.11, 0.1, 0.09, 0.04\\
\end{tabularx}
\caption{A table showing the layer heights and relative strenghts used
  for multiple layer simulations}
\label{table:atmos}
\end{figure}

\subsection{Reconstructor representation}
The control matrix for a single EAGLE science path is likely to
contain of order $10^9$ elements, and must be accessed at a rate of
250~Hz, requiring a memory bandwidth of 1~TBs$^{-1}$ assuming 32-bit
floating point format storage.  When considering that EAGLE is likely
to have up to 20 science paths, the memory bandwidth requirement
increases by a factor equal to the number of science paths, up to
20~TBs$^{-1}$ for EAGLE.

Reducing this memory bandwidth requirement is important to reduce the
real-time control system complexity.  Assuming a \fpga based
wavefront reconstruction unit, the memory bandwidth will be determined
by the \fpga clock rate, the memory to \fpga bus width and the number
of \fpgas used for processing.  By reducing the total size of the
control matrix, the number of \fpgas can be reduced leading to a
cheaper, simpler, more reliable design.  We now consider several
techniques that can be applied to reduce the control matrix size.

\subsubsection{Sparse representation}
\begin{figure}
\includegraphics[width=7cm]{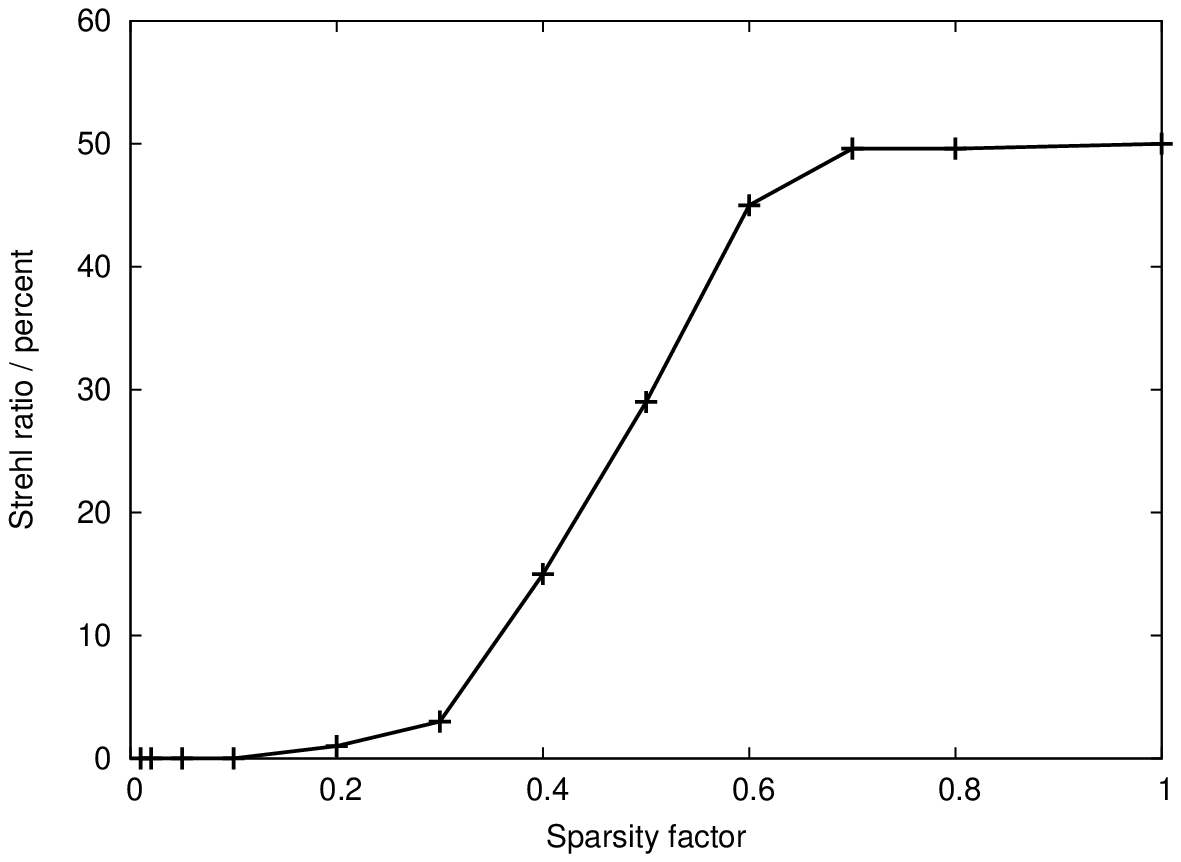}
\caption{A figure showing predicted EAGLE AO system performance as the
  sparsity of the control matrix is altered with the sparsity factor
  representing the fraction of the original control matrix present.
  Uncertainties are about two percent in Strehl ratio.  sparseF2.eps}
\label{fig:sparse}
\end{figure}

Sparse matrix representation of \ao system control matrices has been
studied \cite{1995OptL...20..955W}, and for multi-conjugate systems
(or most systems without a specific \wfs to \dm alignment), sparse
matrix techniques are known to perform poorly \cite{simScaling} due to
poorly sensed modes and \lgs tip-tilt uncertainty.
Fig.~\ref{fig:sparse} verifies this, showing that a highly non-sparse
representation is required to maintain the \ao system performance.
The sparse matrices used here are created by removing the least
influential parts of the control matrix, i.e.\ elements closest to zero.  Typically, 70~\% of the
original matrix must be present, as demonstrated in
Fig.~\ref{fig:sparse}.  However, when stored in sparse format, each
matrix element must be stored accompanied by its position, resulting
in twice as much storage (assuming 32 bit floating point for the
matrix element, and a 32 bit integer for position), thus consuming
more memory than the original control matrix.  Therefore, sparse
matrix representation is not a solution for EAGLE.

\subsubsection{Fixed point representation}
\begin{figure}
\includegraphics[width=7cm]{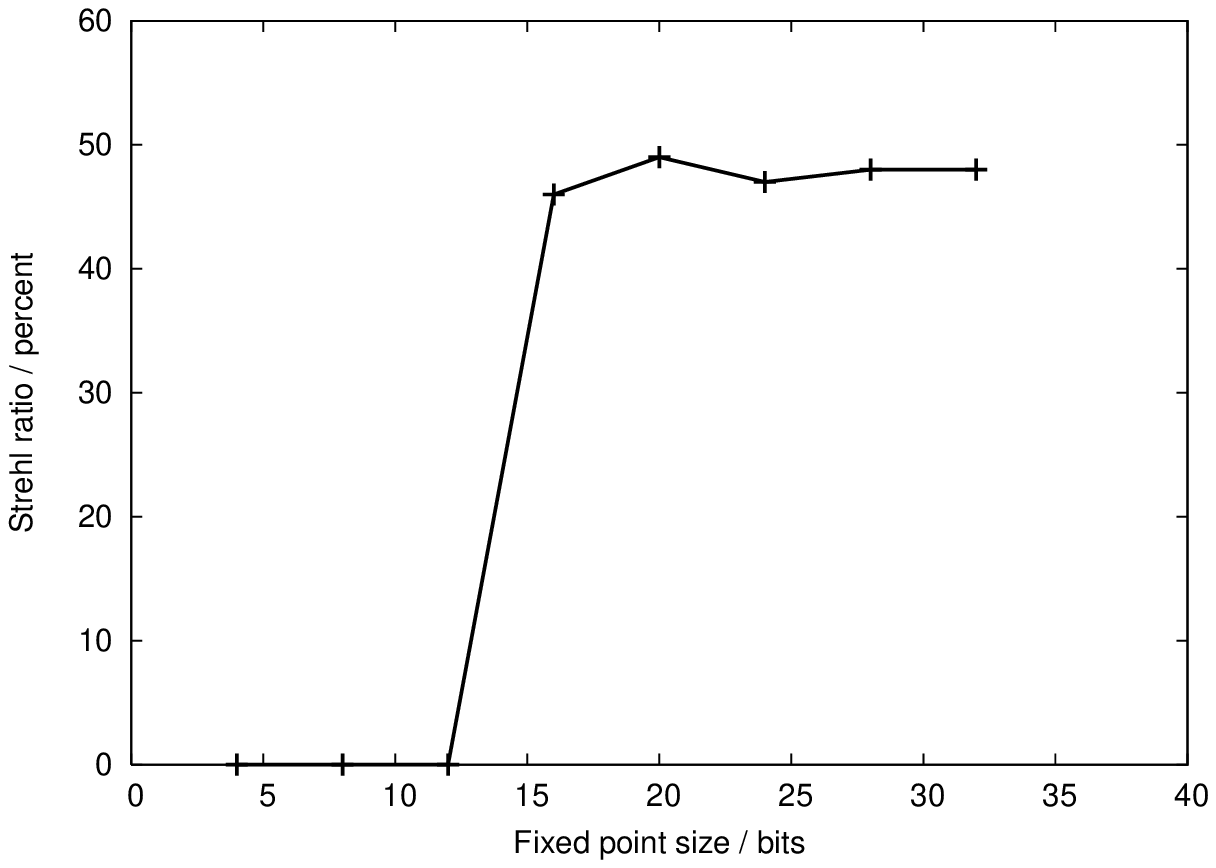}
\caption{A figure showing predicted EAGLE AO system performance with a
  fixed point control matrix representation.  Uncertainties are about
  two percent in Strehl ratio. fixedF3.eps}
\label{fig:fixed}
\end{figure}

Fixed point representation is often used in hardware (for example
\fpgas) as it is simpler to use than floating point representations,
and has a lower computational complexity.  Here, values are stored in
twos-compliment integer format, with a known scaling factor.  By using
a fixed point control matrix representation, as shown in
Fig.~\ref{fig:fixed}, it is possible to reduce the control matrix
storage requirements by a factor of two, using 16 bit fixed point
values rather than 32 bit floating point values, while still
maintaining the \ao system performance.  To compute the fixed point
control matrix, the minimum and maximum elements were found, and used
to compute an offset (equal to the minimum value) and scaling factor
(equal to the range), unique for a given control matrix.  The fixed
point control matrix elements are computed by subtracting the offset
and dividing by the range before being scaled by $2^b$ where $b$ is
the number of bits used to store the fixed point representation.

\subsubsection{Compressed floating point format}
A control matrix is far from homogeneous, with a large range of
values.  This suggests that fixed point representation may not be
ideal, since the relative resolution of small values is low and so
will influence the wavefront error to a greater extent.  We therefore
consider a compressed floating point representation, which is able to
cover the full range of 32 bit floating point, but with a reduced
precision.  Standard IEEE 32-bit floating point values have 8 bits
dedicated to the exponent, 23 bits dedicated to the mantissa and a
single sign bit.  A compressed floating point format which reduces the
precision of the mantissa can be investigated.  In \ao, wavefront
slope measurements are commonly computed using a centre of gravity
measurement, which in good conditions (high light level, low detector
noise) is typically assumed accurate to at best a one hundredth of a
pixel, and in practise, is far less accurate.  With say $20\times20$
pixels per sub-aperture, we can assume that there are 2000 measurable
spot positions across the sub-aperture, which can be encoded in
eleven bits.  Therefore we can predict that a mantissa of a
compressed floating point number need be no more than eleven bits
wide.

\begin{figure}
\includegraphics[width=7cm]{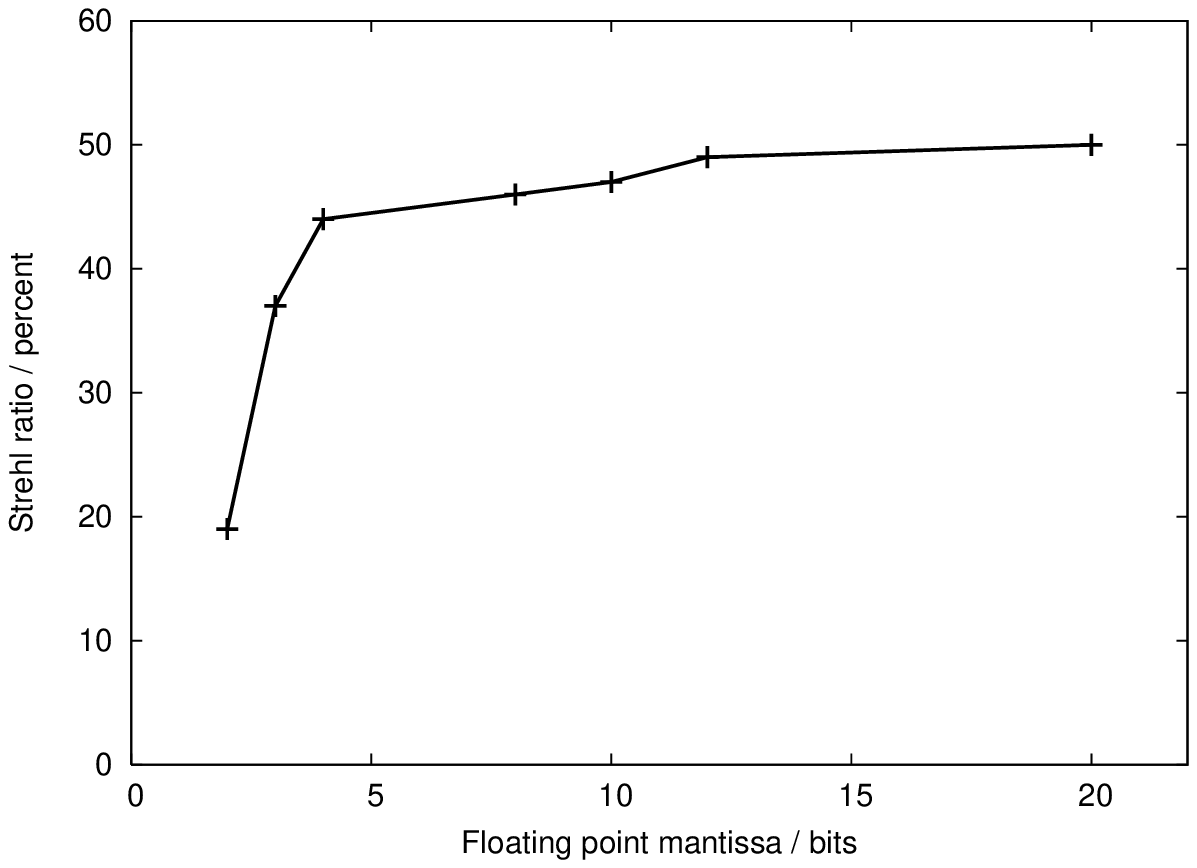}
\caption{A figure showing predicted EAGLE AO system performance with a
  compressed floating point control matrix representation.
  Uncertainties are about two percent in Strehl ratio. comfpF4.eps}
\label{fig:comfp}
\end{figure}

By running \ao simulations with a range of bit-widths for the
mantissa, we find (Fig.~\ref{fig:comfp}) that \ao system performance
is not degraded until fewer than 10--12 bits are used for the
mantissa, which is represented by a compressed floating point number
requiring between 19-21 bits in total.  However, this is a greater
storage requirement than we have shown to be required using a fixed
point representation.

\subsubsection{Variable precision floating point format}
By compressing the exponent, as well as the mantissa of a floating
point number, we can further reduce the storage requirement for the control
matrix.  We represent a floating point number in the form
\begin{equation}
(-1)^s \times b \times a^e \times \left( \frac{a}{2} + m
\right)
\end{equation}
where $s$ is the sign (one bit), $a$ is the base (2 for standard
floating point representation) which is constant for a given control
matrix, $b$ is a scaling factor (constant for a given control
matrix), $e$ is the exponent value, and $m$ is the stored mantissa
value.  As with standard floating point representation, the mantissa
is stored without an implicit integer part, which can be assumed (if
it was not there, the exponent value can always be changed to shift
the mantissa), and this is represented in the equation by the addition
of the mantissa (fixed point with a value less than $\frac{a}{2}$)
with $\frac{a}{2}$.  The exponent, $e$ is in standard twos-compliment
integer format.

To convert a standard control matrix into this format, the minimum and
maximum values required for storage are first obtained.  We then set
requirements that the mantissa for the maximum value is all ones, and
the mantissa for the minimum (non-zero) value is all zeros except for
the final bit, which is set.  The exponent for the maximum value has
all bits set, and the exponent for the minimum value has all bits
unset.  A value of zero is represented by having all bits of the
mantissa and exponent unset.  These conditions allow us to find two
unknown values, $a$ and $b$ which will allow us to store this control
matrix with highest precision.  We then proceed to convert the
standard control matrix into the variable precision representation.

\begin{figure}
\includegraphics[width=7cm]{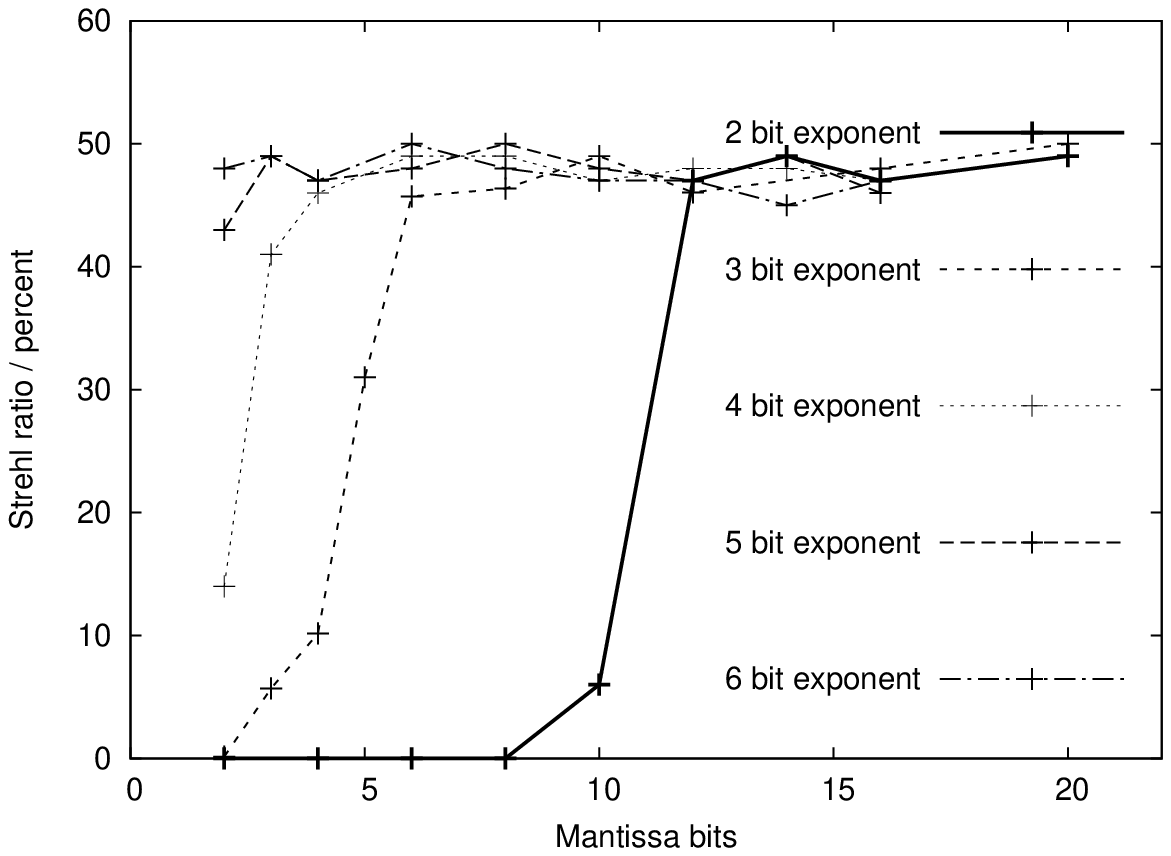}
\caption{A figure showing predicted EAGLE AO system performance with a
  variable precision floating point control matrix representation.
  The key gives the number of bits used for the exponent for each
  curve.  Uncertainties are about two percent in Strehl ratio. varfpF5.eps}
\label{fig:varfp}
\end{figure}

As can be seen from Fig.~\ref{fig:varfp}, using a four bit exponent
and mantissa is sufficient for good \ao system performance, i.e.\ a
total of nine bits per control matrix element (including a sign bit).
Similarly, a five bit exponent and three bit mantissa, and a six bit
exponent and two bit mantissa also provide similar performance (taking
nine bits per control matrix element).  The memory storage requirement
has therefore been reduced by almost a factor of four.  We have not
investigated the effect of using a greater number of atmospheric
layers and virtual \dms, though at most, this will increase the number
of bits required slightly.

By using variable precision floating point for storage of the control
matrix, the memory bandwidth requirement can be reduced by a factor of
almost four, which will greatly simplify the design of a real-time control
system for EAGLE.  Only a quarter of the \fpgas used by an
uncompressed system would be required, with simplifications also made
by reducing the number of inter-\fpga connections, and an increased
reliability due to a reduced number of components.

\subsubsection{Implementation in FPGA}
Implementation of variable precision floating point format in an \fpga
is trivial: A $2^{4+4}=256$ element look-up table can be used to
translate the stored 9 bit control matrix values (using the mantissa
and exponent for the index into the look-up table) into standard
32-bit floating point values to which the sign can then be inserted.
A standard floating point multiplication routine can then be used
during the matrix-vector multiplication.  

The Virtex-6 family of \fpgas is the latest offering from the company
Xilinx, one of the major manufacturers of these devices.  This range
includes devices with up to 1200 input/output pins, with 37~MB
internal memory in the \fpga (similar to a \cpu cache), and a clock
rate of up to 1.6~GHz.  The design of a real-time control system could
be carried out using internal memory only.  In this case, to store 20
control matrices (one for each line-of-sight) each 4~GB in size (1
billion 32-bit floating point values), would require over 2000 \fpgas.
Using variable precision floating point can reduce this requirement to
just over 600 \fpgas, though this is still an undesirably large number.

Alternatively, we can use external memory connected to the \fpga pins.
With standard 32-bit floating point storage, and a 1024 bit wide
memory bus (1024 \fpga pins connected to memory), we can access 32
values each memory read.  The remaining pins are reserved for the
address bus, and inter-\fpga communications.  Assuming that memory can
be accessed at the full \fpga clock rate (1.6~GHz) we will achieve a
memory bandwidth of about $5\times10^{10}$ values per second.  The
requirement for EAGLE is a minimum of $5\times10^{12}$ values per
second, so 100 \fpgas would be required.  If however, variable
precision floating point storage is used, we could access 114 values
each memory read (with a 1026 bit wide \fpga bus), equating to
$1.8\times10^{11}$ values per second, requiring 28 \fpgas for EAGLE, a
far more attractive proposition to develop.

In practice, the memory bandwidth requirement may be increased to
reduce the \ao system latency.  Here, we have assumed a latency of
4~ms, equal to the frame time (at 250~Hz).  To achieve a latency of
1~ms, we would require an improvement in memory bandwidth by a factor
of four, which in turn would require 112 \fpgas to meet this
requirement.

\subsection{Wavefront sensor calibration}
Shack-Hartmann based wavefront sensors are slightly non-linear due to
the pixelated nature of the detector meaning that position information
is lost: The measured slope is not proportional to the actual
wavefront slope across the sub-apertures.  For closed loop \ao
systems, this is not a problem since the degree of non-linearity is
small and because the measured wavefront slopes are minimised by the
\dm, a linearity approximation works well.  However, for typical
open-loop systems, this is more problematic since large uncorrected
wavefront slopes can be measured.  Therefore, the corrected wavefront
(unsensed) will have some additional error due to this non-linearity.
This error is enough to lead to reduced performance of the \ao system,
and is present regardless of the slope measurement algorithm used if
this algorithm is linear (e.g.\ centre of gravity, matched filter and
correlation algorithms).  However, a suitable calibration of the \wfs
can be carried out, measuring the \wfs estimated response to a set of
known incident wavefront slopes (introduced by a flat mirror on a
tip-tilt stage).  During \ao system operation, the uncalibrated
measured wavefront slope can then be used to infer the true
(calibrated) wavefront slope by interpolating from the calibration
data.  This calibrated measurement can then be used to perform a more
accurate wavefront reconstruction.

\begin{figure}
\includegraphics[width=7cm]{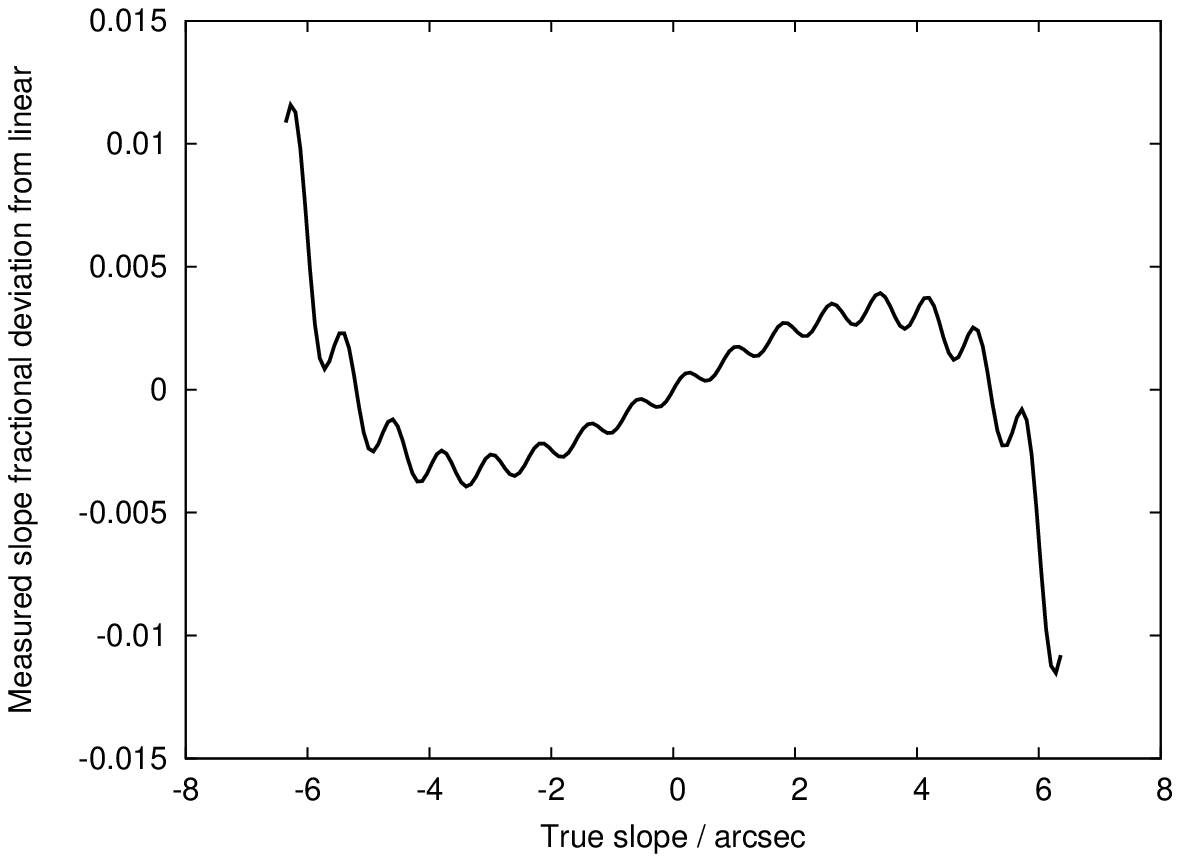}
\caption{A figure showing the non-linearity of a Shack-Hartmann
  wavefront sensor, after subtraction of the linear response. calibrationF6.eps}
\label{fig:calibration}
\end{figure}

\begin{figure}
\includegraphics[width=7cm]{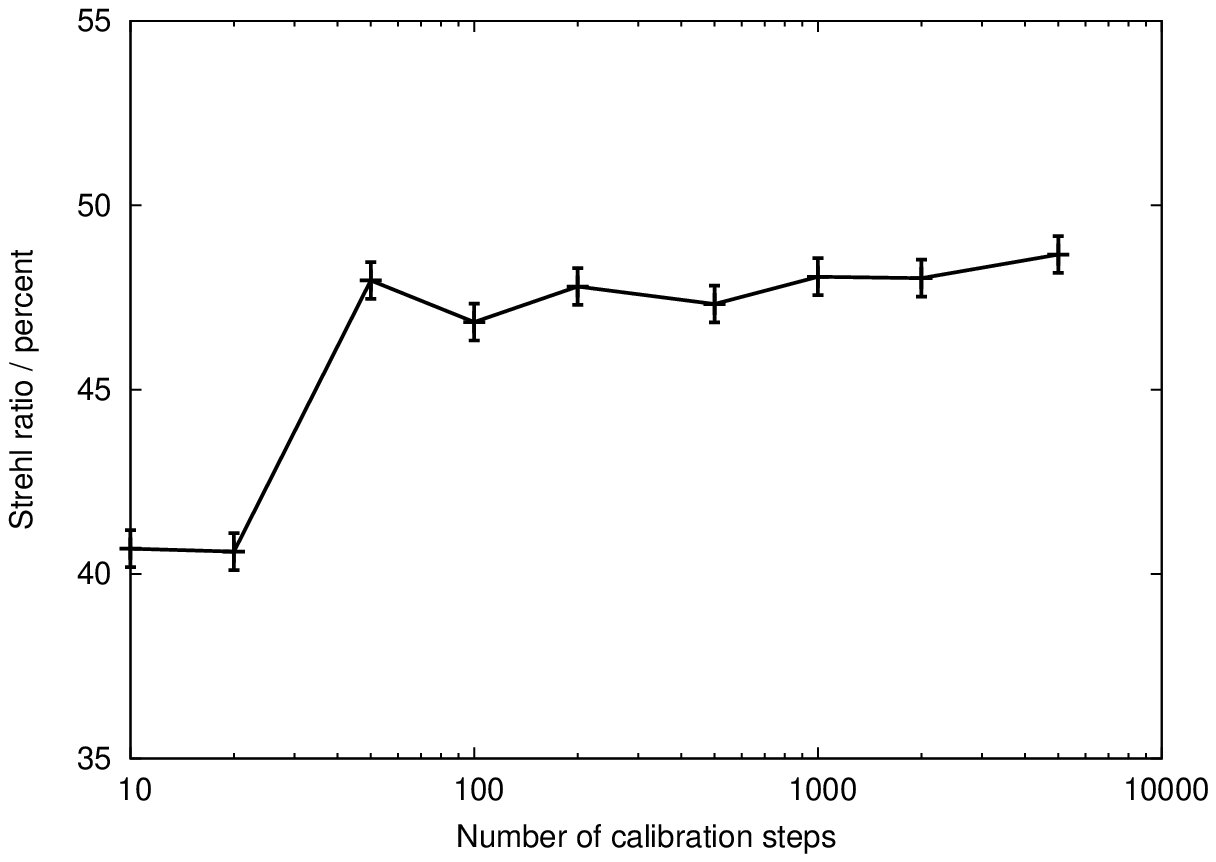}
\caption{A figure showing predicted EAGLE AO system performance
  (Strehl ratio) as the number of calibration steps is increased.  A
  40\% Strehl ratio is achieved with no calibration (0 steps, not
  shown on the logarithmic scale). linearstepsF7.eps}
\label{fig:linearsteps}
\end{figure}

We have performed Monte-Carlo simulations using this technique for
\wfs calibration using a centre-of-gravity slope measurement
algorithm, and have investigated the number of calibration steps
required for good \ao performance.  These simulations are based around
the aforementioned EAGLE simulations.  We have used Shack-Hartmann
sub-apertures with $20\times20$ pixels each, and a pixel scale of
0.8~arcsec per pixel at a wavelength of 589~nm and as before, Fried's
parameter is 10.6~cm.  The \wfs calibration is performed over the
entire sub-aperture field of view.  This large field of view is due to
the need to detect the elongated \lgs spots, and due to the higher
dynamic range of the open-loop \wfs (spots are measured in open-loop,
so are not necessarily close to the centre of sub-apertures, as is
usually the case for a closed loop system).  Fig.~\ref{fig:calibration}
demonstrates the degree of non-linearity in the simulated
Shack-Hartmann wavefront sensor showing the deviation of measured
wavefront slope from the true slope as the true spot position moves
across the sub-aperture.  A true (physical) sensor would display even
more non-linearity due to imperfect optics.
Fig.~\ref{fig:linearsteps} shows the performance improvements achieved
with increasing calibration accuracy (number of calibration steps),
demonstrating that this linearisation calibration is an important part
of open-loop \ao system operation.  We see that in this case, at least
50 slope calibration measurements are required to achieve best
performance, each step corresponding to a spot shift of less than half
a pixel.  By performing this calibration, the Strehl ratio (relative
to uncalibrated performance) is increased by over 25\%, and so the
design of an open-loop real-time control system should therefore
incorporate this calibration step.  It should be noted that the
optimal number of calibration steps is dependent on the \wfs spot size
on the detector so will vary with instrument and atmospheric
conditions.


\subsection{TMT comparisons}
The \tmt project also has plans for a multi-object spectrograph with
\ao, IRMOS \cite{2006SPIE.6272E..23G,2006SPIE.6269E.145A}.  The
results presented here show that the estimated performance of these
systems (taking into account the many unknowns in the designs), both
estimating 50--60\% ensquared energy in 50~mas.  It should be noted
that the results presented in this paper have been for energy within
75~mas.  When we use our simulation models to measure energy within
50~mas, this is typically about 1--2\% lower than the energy within
75~mas.  This serves to strengthen the assumption that modelling of
\ao systems can yield reliable performance estimates.

\section{Conclusion}
We have performed full end-to-end Monte-Carlo simulations of an \ao
system for EAGLE.  Investigations reported here show that the
atmospheric turbulence profile must be well known, with the heights of
turbulent layers known to within a few hundred metres.  We have also
reported on an investigation of compressed reconstructor
representations and find that it is possible to reduce control matrix
memory requirements by almost a factor of four in the cases
investigated, significantly reducing the complexity of an \fpga based
real-time control system.  An investigation into the effect that the
non-linearity of Shack-Hartmann based wavefront sensors has on \ao
system performance has also been carried out, demonstrating that a
linearity calibration should be included in an open-loop real-time
control system to improve performance.

\section*{Acknowledgements}
This work is funded by the STFC.


\end{document}